# Influence of acoustic waves on Radiation Spectra of Argon Gas-discharge Plasma

## Aramyan A.R.

Institute of Applied Problems of Physics, NAS RA
25, Hr. Nersessian Street, Yerevan, 375014, Republic of Armenia
E-mail: ara@iapp.sci.am

It is shown that under the definite regime of interaction of the acoustic waves with low-temperature, partially ionized plasma it is possible to change the intensity of some spectral lines of atomic argon. It is shown also, that the dependence of the intensities of these spectral lines on the intensity of the acoustic wave has a hysteresis behavior.

The research on properties of low-temperature ionized plasma and on the influences on it is always urgent, since besides the elucidation of new physical mechanisms, the results of these investigations quickly find practical application. It is noteworthy, however, that the influence of acoustic waves on plasma parameters and on processes in the plasma environment are studied relatively poor. Besides, high interest arose lately to the problem of interactions of acoustic waves with such thermodynamically nonequilibrium gas as is the partially ionized plasma of gas discharge, where the electron temperature usually much exceeds that of heavier particles [1–6]. Note that of numerous problems connected with the interactions of acoustic waves with partially ionized plasma, the special promising is the influence of acoustic waves on radiation spectra of gas-discharge plasma. Last years, an interesting effect consisting in sharp change of radiation spectrum under action of sonic wave in dense (p ~ 100 Torr) gas discharge plasma of argon was observed in our laboratory [7]. After several seconds from cutting out the sonic wave, in different points of the discharge tube in the bulk of positive column flashes of radiation were observed during several minutes, that corresponded to definite transitions between electronic levels of atomic argon. An assumption was made about the autogeneration of appropriate lines in the spectrum of argon.

The studies of the change of a radiation spectrum under the influence of acoustic waves were conducted using low-temperature argon gas discharge plasma (pressure – 100 Torr, discharge current - 50 mA, voltage on electrodes 2 kV). The experimental set-up (Fig.1) is a quartz discharge tube with internal diameter of 60 and length of 1000 mm. The distance between the electrodes was 850 mm. An electrodynamic transmitter is attached to one of butt ends of the

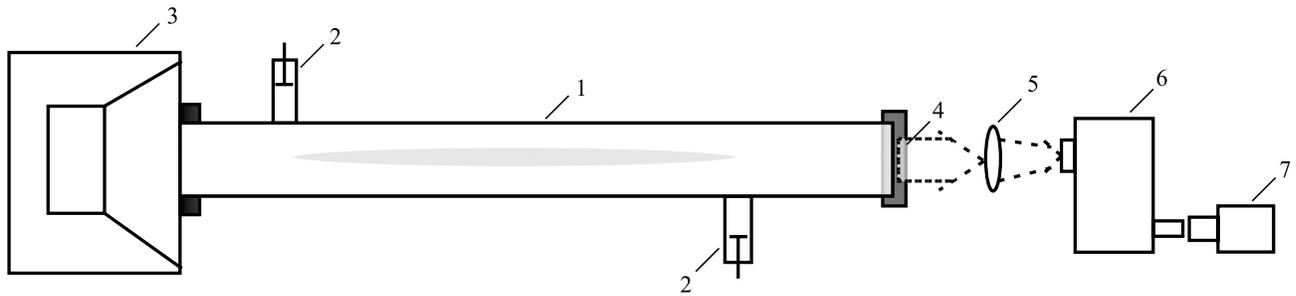

Figure1. The experimental set up: 1 – is the discharge tube; 2 – electrodes; 3 – the electrodynamic transmitter of sonic waves; 4 – the window; 5 - the lens; 6 – the monochromator; 7 – the photoelectric multiplier tube.

tube. The light emitted from plasma is extracted through the second butt to a monochromator that is used for investigation of variations of some spectral lines of plasma radiation under the action of acoustic waves.

The investigations showed [7] that variations of argon plasma radiation are manifested as flashes with duration of 15 – 20 ms. As is seen from Fig.2 c, d, the flashes are of orange and blue colour. These flashes arise independently one of the other. The orange flashes appear on the outer side of the border of discharge pinch, whereas the blue flashes burst inside the discharge pinch.

An analysis of the spectral composition of flashing radiation showed that these correspond to three transitions between energy levels of atomic argon:

$7p'[1/2]_1$ - $4p\,[1/2]_1$ wave length 4876 Å

$7s\,[3/2]_1$ - $4p\,[5/2]_3$  wave length 5888 Å

$4d'[3/2]_1$ - $4p\,[1/2]_1$  wave length 5882 Å

A diagram of the energy levels of these transition in argon is presented in Fig. 3. As one can see from this figure, the oscillator strengths of the 4p-4s transition are approximately 30-100 times greater than the oscillator strengths of the above-indicated transitions. For this reason, the 4p level is rapidly depopulated. As a results, a population inversion can arise between each of the levels 7p', 7s and 4d' on the one hand and the 4p level on the other. This population inversion can evidently serve at the reason for the appearance of the observed flashes. As a result of further research of these lines it was shown that under definite interaction regime of acoustic waves with plasma the change in the plasma radiation is manifested not only by flashes, but also by essential amplification of above spectral lines.

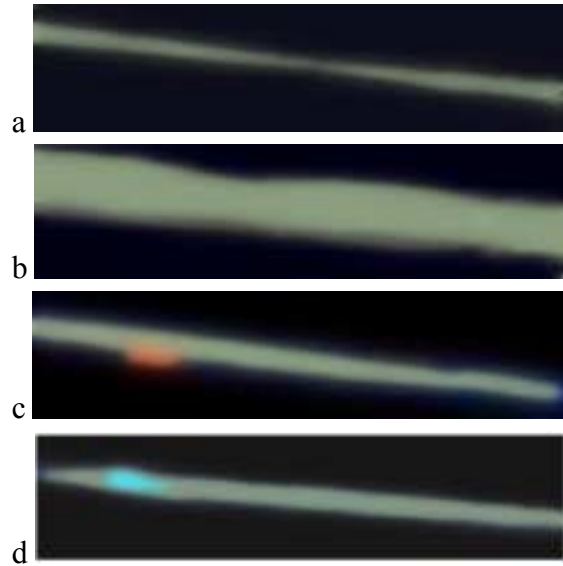

Figure 2. Visual change of pinched-plasma column under the influence of resonant (190 Hrz) acoustic waves: a)   the pinched-plasma column in the absence of acoustic waves; b)   the plasma under the influence of intense (of more  than 82 dB) resonant sonic waves; c,d) the plasma under the influence of resonant sonic waves (of intensity less than 82 dB).

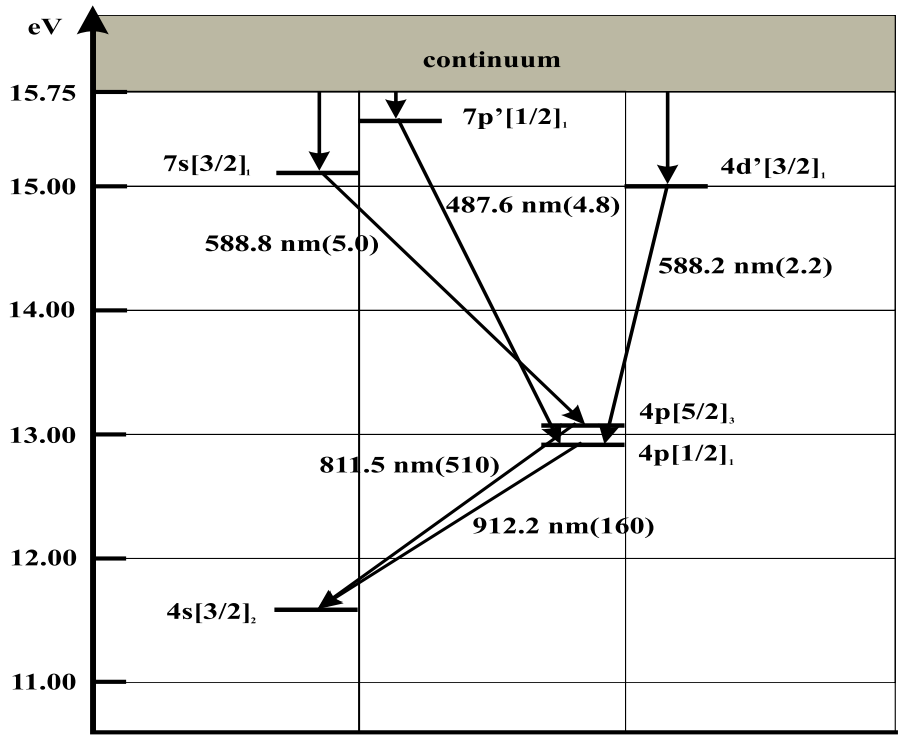

Figure 3. The diagram of energy levels of some transitions in atomic argon.

The spectra of argon plasma radiation taken with the help of the aforementioned device are shown in Fig.4: the spectrum of plasma radiation prior to the interaction with acoustic waves is given in Fig.4a; the spectrum of plasma radiation under the influence of intense acoustic waves (of more than 82 dB) and of acoustic waves with intensity less than 82 dB are shown in Figs.4b and figure 4(c) respectively.

The dependence of radiation intensity of $4d'[3/2]_1$-$4p[1/2]_1$ transition on the acoustic wave intensity at frequency of 190 Hertz is given in Fig.5 and is seen to have hysteresis behavior. When the intensity of acoustic waves increases (from the zeroth value) to $A_{max}$ (that corresponds to 90 dB), no any change in the radiation spectrum is observed and for these intensities of acoustic waves a decontraction of plasma takes place and it fills the volume of tube completely (Fig.2(b)). During the return path of acoustic wave intensity, a notable increase in line radiation intensity is observed at a definite value $A_c$. The critical value $A_c$ corresponds to the unpinching threshold of the discharge (82 dB). The line intensity is observed to increase after the pinching of discharge. A further lessening of acoustic wave intensity is accompanied with smooth decrease of the line intensity

For determination of relation between the constant increase of spectral line intensity and the flashes we have plotted the dependences that will be discussed below. At the return path A, the reduction in acoustic wave intensity stays on the value $A_o$, that corresponds to 80 dB and is less than $A_c$ (Fig.5). In figure 6(a) the time dependence of radiation intensity of $4d'[3/2]_1$-$4p[1/2]_1$ transition under constant influence of resonant acoustic waves with frequency of 190 Hertz and intensity $A_0$ is presented. It is seen, that in the absence of flashes the intensity of line radiation has a constant value $I_o$. During the bursting of flashes the line intensity sharply increases (in nearly 100 times) up to $I_{max}$. After the bursting of a flash (with duration of 15 – 20 ms) the intensity drops to the minimum $I_{min}$, which corresponds to the value of radiation intensity in the absence of acoustic waves.

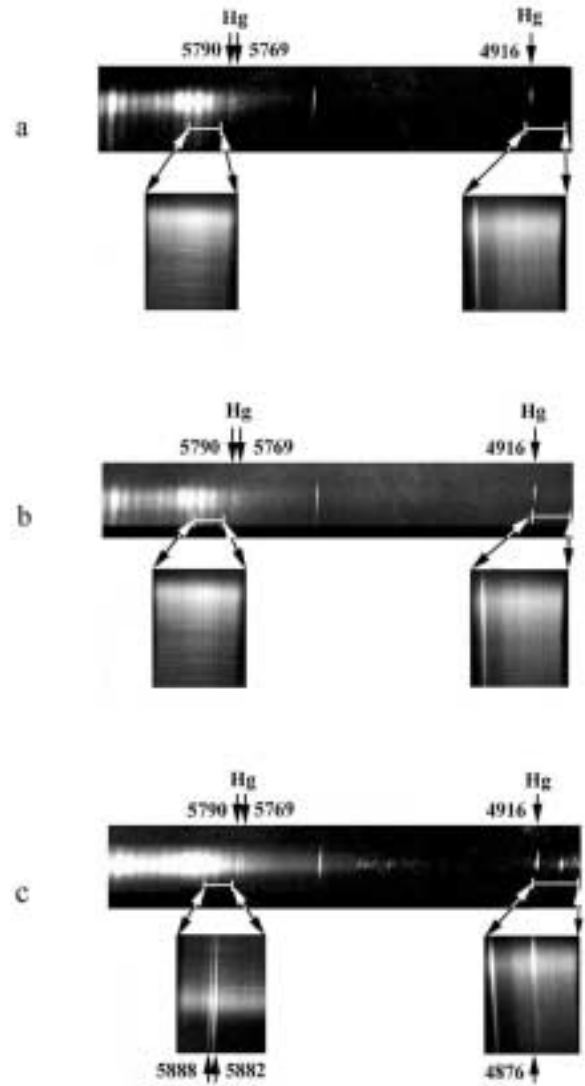

Figure 4. The spectra of argon plasma radiation: a) the spectrum of plasma radiation prior to the interaction with acoustic waves: b) the spectrum of plasma radiation under the influence of intense acoustic waves (of more than 82 dB): c) the spectrum of plasma radiation under the influence of acoustic waves (of less than 82 dB).

The rise of intensity from $I_{min}$ to $I_o$ is rather long, ~1s. Shown in the second plot (Fig.6(b)) is the dependence of radiation intensity of an analogous transition after cutting-out of acoustic waves, that has the initial frequency of 190 Hertz and intensity – 90 dB. After cutting-out of acoustic waves the line intensity keeps on at the minimum initial value of $I_{min}$ during several seconds (~2s). The value of $I_o$ is reached in $t_o$ time and at the further increase in time, $I_o$ smoothly decreases to $I_{min}$ in 15 – 20 s. The pattern of bursting of flashes in this regime is analogous to the previous case (Fig.6(a)).

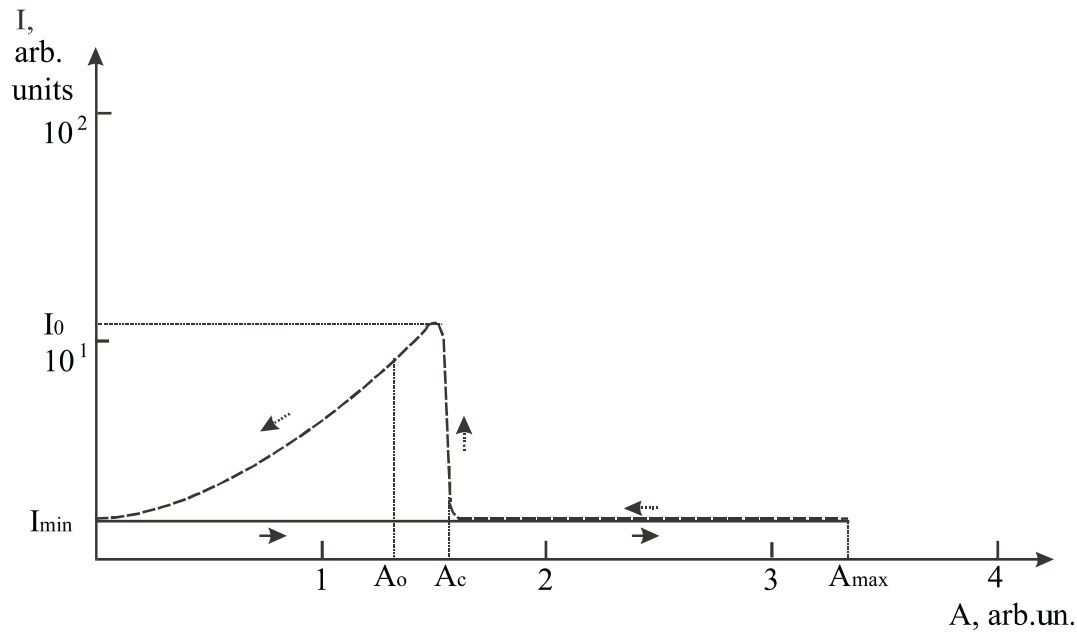

Fig.5 Dependence of the radiation intensity of 4d'[3/2]$_1$ - 4p[1/2]$_1$ transition on the acoustic wave intensity at the frequency of 190 Hz.

For explanation of this effect recall that the sonic wave can strongly change the parameters of plasma [8] including the concentration and temperature of electrons $N_e$ and $T_e$. It is known that an intense resonant acoustic wave gives rise to vortex motion of gas [9] in a volume(in the presence of a gradient of oscillatory velocity). These vortex motions of the gas are known as "acoustical flows". The indispensable condition for the rise of such flows in cylindrical vessels (similar to our discharge tube) is $\lambda \gg R$, where $\lambda$ is the wavelength and R is the radius of tube. The diagram of such flows in a tube is shown in Fig.7. The velocity of acoustic flow depends on the acoustic wave intensity, and changes of flow velocity may lead to a change of $N_e$ and $T_e$ in different points. For cases realized in our experiments, the change of $N_e$ and $T_e$ is pronounced, e.g., in places, where the enhancement of 4d'[3/2]$_1$ - 4p[1/2]$_1$ line is observed (outside the radial border of pinch). The population of highly excited levels of argon atoms in these thermodynamically manifest nonequilibrium ranges may strongly increase and there exists a possibility of inversion between definite levels. As is known, there are two types of nonequilibrium,- the ionization and the recombination ones. In the first case the local degree of ionization is less than the equilibrium one, corresponding to $T_e$, and in the second case, on the contrary, is more. At a definite value of acoustic flow intensity in places, where the enhancement of 4d'[3/2]$_1$ - 4p[1/2]$_1$ line is observed, the recombination nonequilibrium is effected. Most probably, the dependence of acoustic flow intensity in the discharge tube on acoustic wave intensity has a hysteresis behavior. The required intensity of acoustic flow is obtained on the return path of hysteresis loop, i.e., at the reduction of acoustic wave intensity.

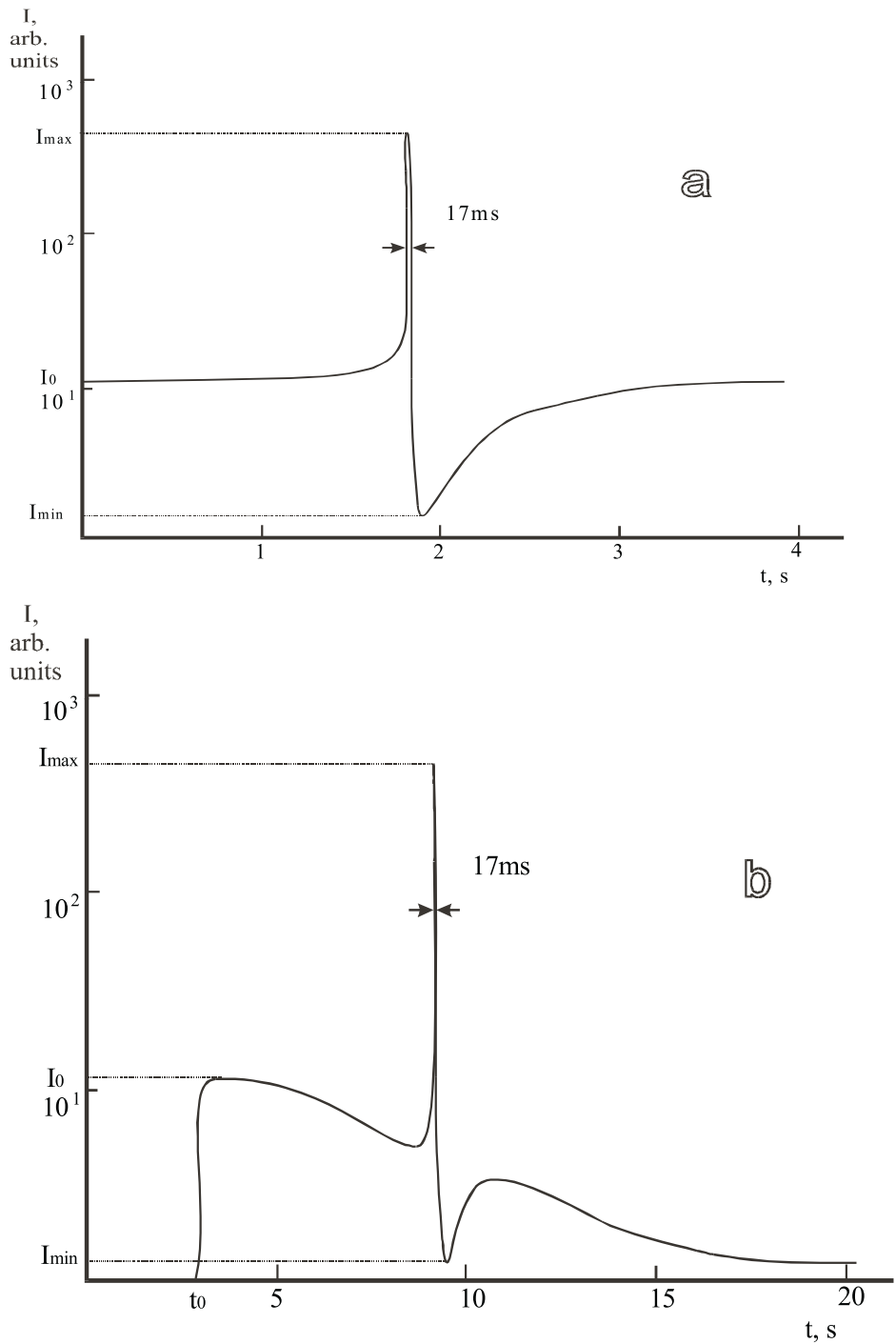

Fig.6  The time dependence of the radiation intensity of 4d'[3/2]$_1$ - 4p[1/2]$_1$ transition: a) under constant action of resonant acoustic waves of frequency 190 Hz and intensity 80 dB; b) after cutting–off of acoustic waves, the initial value of frequency of which was 190 Hz and intensity–90 dB.

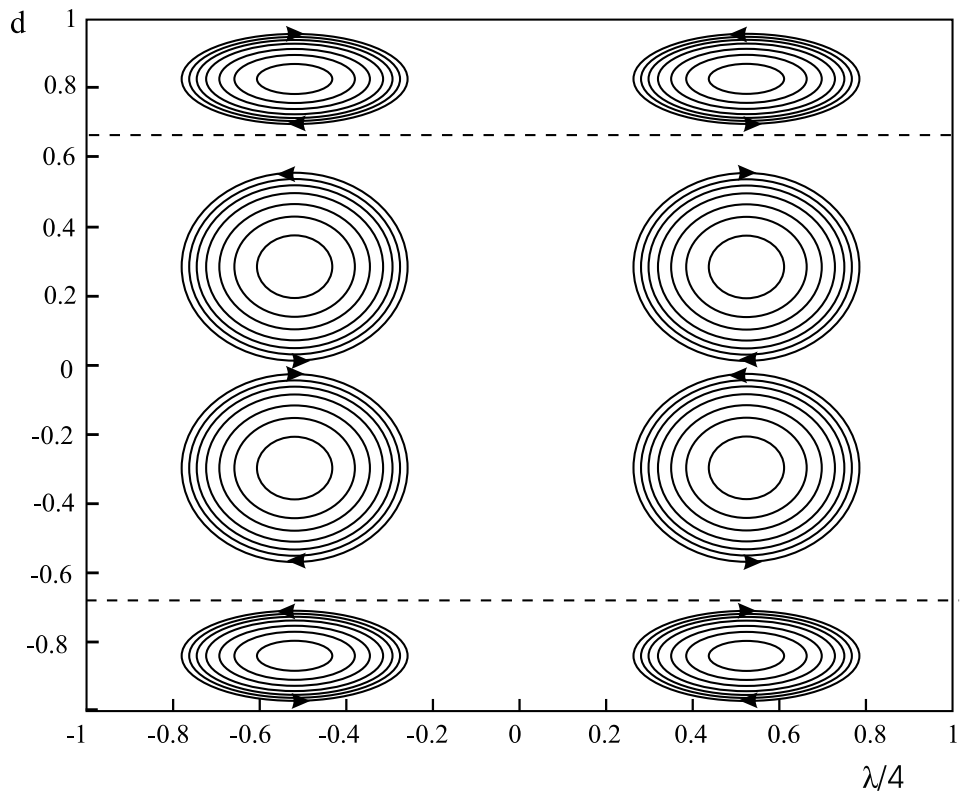

Fig.7. The diagram of "acoustical flows" in a tube.

I am deeply grateful to Academician A.R. Mkrtchyan for his faithful attention, consultations and assistance in this work.